\documentclass[oneside,a4paper,12pt,shownumbers,manualsort]{article}
\usepackage{latexsym}
\usepackage{euscript}
\usepackage{epsfig,amsmath,amssymb}

\renewcommand\({\left(}
\renewcommand\){\right)}
\renewcommand\[{\left[}
\renewcommand\]{\right]}

\newcommand{\e}{{\rm e}}
\def\be{\begin{equation}}
\def\ee{\end{equation}}
\def\bea{\begin{eqnarray}}
\def\eea{\end{eqnarray}}

\newcommand\GeV{\,\mbox{GeV}}

\newcommand\eV{\,\mbox{eV}}
\newcommand\mpl{m_{\rm p}}
\newcommand\mcN{\mathcal N}

\long\def\symbolfootnote[#1]#2{\begingroup%
\def\thefootnote{\fnsymbol{footnote}}\footnote[#1]{#2}\endgroup}

 \large\normalsize

\begin{document}

\begin{center}
  {\Large \bf Leptogenesis from reheating after inflation and cosmic
  string decay}

\vspace*{7mm}
{\ Rachel Jeannerot$^{a}$\symbolfootnote[1]{{E-mail:jeannerot@lorentz.leidenuniv.nl}} and Marieke Postma$^{b}$}\symbolfootnote[2]{E-mail:mpostma@nikhef.nl}
\vspace*{.25cm}

${}^{a)}${\it Instituut-Lorentz for Theoretical Physics,
Niels Bohrweg 2, 2333 CA Leiden, The Netherlands}\\
\vspace*{.1cm} 
${}^{b)}${\it NIKHEF, Kruislaan 409, 1098 SJ Amsterdam,
The Netherlands}

\begin{abstract}

Cosmic strings form at the end of standard supersymmetric hybrid
inflation, and both inflation and strings contribute to the CMB
anisotropies. If the symmetry which is broken at the end of inflation
is gauged $B$-$L$, there is a mixed scenario for leptogenesis:
Right-handed neutrinos can be produced non-thermally during reheating
via inflaton decay as well as via cosmic string decay. We show that
the parameter space consistent with CMB data can accommodate either or
both scenarios depending on the mass of the right-handed neutrinos.

\end{abstract}
\end{center}


\section{Introduction}

Supersymmetric theories beyond the standard model which predict light
neutrino masses via the see-saw mechanism easily accommodate SUSY
hybrid inflation \cite{Cop,Dvasha}. Reheating can proceed via inflaton
decay into right-handed (s)neutrinos and thereby these models also
provide an interesting framework for non-thermal leptogenesis
\cite{Yanagida,leptinfl,shafi1}. In the simplest version of hybrid
inflation cosmic strings form at the end \cite{prd,jrs}; if the
inflaton sector couples to right-handed (RH) (s)neutrinos these are
$B$-$L$ cosmic strings \cite{lept,B-L}. $B$-$L$ cosmic strings are not
superconducting \cite{maj,sugrastr}. Most of the energy lost by the
string network goes into gravitational radiation and right-handed
neutrinos, and therefore these strings provide a second mechanism of
non-thermal leptogenesis \cite{lept}. In this paper we calculate the
relative contributions to the baryon asymmetry of the universe from
reheating at the end of standard $F$-term inflation and from $B$-$L$
cosmic string decay.

Let $G_{\rm GUT}$ denote a gauge group which contains the Standard
Model gauge group as well as gauged $B$-$L$. $F$-term inflation
requires the existence of a gauge singlet and two Higgs superfields
which transform in complex conjugate representations of $G_{\rm GUT}$;
we assume that they break $B$-$L$ when acquiring a non vanishing
vacuum expectation value (VEV). In the case of SO(10), they could be a
$16 + \overline{16}$ or $126+\overline{126}$. Note that the
transformation properties of the Higgs representation can affect the
stability of the strings; we shall not discuss it further here
\cite{B-L}.  Inflation takes place as the scalar singlet slowly rolls
down a valley of local minima along which the VEV of the Higgs fields
vanish. When the singlet falls below a certain critical value, the
Higgs mass become tachyonic and inflation ends quickly in a phase
transition during which the Higgs fields acquire a non-vanishing VEV
breaking $B$-$L$ spontaneously. If $G_{\rm GUT}$ is semi-simple, the
assumption of standard SUSY hybrid inflation then requires that
$G_{\rm GUT}$ breaks down to the Standard Model (SM) gauge group via
at least one intermediate step, so that inflation solves the monopole
problem; the gauge group broken at the end of inflation is not $G_{\rm
GUT}$ but an intermediate symmetry group $G_{\rm int} \supset
U(1)_{B-L} \rightarrow H \not\supset U(1)_{B-L}$. During this phase
transition, $B$-$L$ cosmic strings form. The simplest example is
$G_{\rm GUT} = SO(10), E(6)$ or Pati-Salam, and $G_{\rm int} = SU(3)_c
\times SU(2)_L \times U(1)_R \times U(1)_{B-L}$ which breaks down to
the SM gauge group at the end of inflation \cite{prd1,prd2}.

The GUT Higgs fields which trigger the end of inflation give heavy
Majorana masses to the right-handed neutrinos. After inflation the
universe is reheated by inflaton decay into RH neutrinos and
sneutrinos.  If the reheat temperature is less than the neutrino mass
their out-of-equilibrium decay into (s)leptons and the SM Higgs(inos)
produces a net lepton asymmetry \cite{leptinfl}.  There is another
contribution to the lepton asymmetry, coming from the decay of cosmic
string loops~\cite{lept}. Cosmic string loops decay into $B$-$L$ Higgs
and gauge fields, which in turn decay in right-handed neutrinos. Loops
can also release right-handed neutrinos which are trapped as zero
modes.

Thus both inflation and cosmic strings contribute non-thermally to the
baryon asymmetry of the universe. We investigate here which of these
scenarios is most efficient using the observed CMB anisotropies as a
constraint.  The string tension, as well as the inflaton mass are set
by the symmetry breaking scale at the end of inflation $\eta$ (the
$B$-$L$ breaking scale) and the Higgs self quartic coupling $\kappa$
(the superpotential coupling). The string tension also depends
logarithmically on the gauge coupling constant which we set to the
unification value for the MSSM. The resulting lepton asymmetry depends
on $\eta$, $\kappa$ and the RH neutrino masses. Requiring that the
inflaton gives the observed density perturbations fixes $\eta$ as a
function of $\kappa$.  The string contribution to the density
perturbations is constrained to be less than about
10\%~\cite{Pogosian}. This restricts the value of $\kappa$, as
discussed in our previous paper \cite{CMB}.

Most of the lepton asymmetry in $B$-$L$ string decay is generated at
the earliest time at which leptogenesis is possible.
Refs. \cite{Sahu,Gu} assume that the reheat temperature is high enough
for the lightest RH neutrino with mass $M_1$ to be initially in
thermal equilibrium.  Any asymmetry generated is washed-out until the
lightest RH neutrino freezes out.  Hence, in their analysis the lepton
asymmetry is dominated by the contribution generated at the freeze-out
temperature $T \sim M_1$. However, there is also the possibility that
the RH neutrinos are never in thermal equilibrium after
inflation. Then the loops formed immediately at the end of inflation
give the dominant contribution to the lepton asymmetry.  The final
asymmetry is determined by three factors.  First, it depends on the
initial string density, which can be different from the density during
the scaling regime.  Second, the universe goes from matter domination
to radiation domination during reheating.  The earlier this happens,
the larger the asymmetry. And the third factor which plays a r\^ole is
the CP asymmetry per decaying (s)neutrino, which depends on the
details of the neutrino sector.

The paper is organised as follows. In Sec.~2 we review standard SUSY
$F$-inflation coupled to N=1 SUGRA and discuss the CMB constraints
\cite{CMB}. In Sec.~3 we determine the parameter space for successful
non-thermal leptogenesis from reheating at the end of inflation. In
Sec.~4 we turn to the parameter space for successful non-thermal
leptogenesis which results from the decay of $B$-$L$ cosmic
strings. We distinguish three different cases. The first possibility
is that $M_1 < m_\chi/2$ and $M_1>T_{\rm R}$, with $M_1$ the lightest
RH neutrino mass, $m_\chi$ the inflaton mass and $T_{\rm R}$ the
reheat temperature.  In this case reheating goes via production of RH
neutrinos which are out-of-equilibrium at the end of inflation, and
both non-thermal leptogenesis scenarios compete.  The second
possibility is that $M_1 > m_\chi/2$ and reheating is
gravitational. Then cosmic strings give the sole contribution to the
lepton asymmetry. Finally there is the possibility that $M_1<T_{\rm
R}$ and the RH neutrinos are in thermal equilibrium at production. In
this case both cosmic string decay and standard thermal leptogenesis
contribute to the lepton asymmetry.  In Sec.~5 we consider the
consequences on leptogenesis of generating a dynamical $\mu$-term.
We give our conclusions in Sec. 6.

\section{Hybrid inflation \& CMB constraints}
\label{s:HI}

In this section we summarise the bounds on the parameter space implied
by current CMB data for standard hybrid inflation with cosmic
strings. The details can be found in Ref. \cite{CMB}.

The superpotential for standard hybrid inflation is 
\cite{Cop,Dvasha}
\be 
W_{\rm inf} = \kappa S( \phi \bar{\phi} - \eta^2),
\label{W}
\ee
with $S$ a gauge singlet superfield, and $\phi$, $\bar{\phi}$ Higgs
superfields in $\mcN$-dimensional complex conjugate representations of
a gauge group $G_{\rm GUT}$.  Upon acquiring a VEV the Higgs fields
break $G_{\rm int} \supset U(1)_{B-L} $ down to a subgroup $H
\not\supset U(1)_{B-L}$.  The supersymmetric part of the scalar
potential is given by (we represent the scalar components with the
same symbols as the superfields)
\be
V_{\rm SUSY} = \kappa^2 |\phi \bar{\phi} - \eta^2|^2 + \kappa^2
|S|^2(|\phi|^2+|\bar{\phi}|^2) + V_D.
\label{Vsusy}
\ee
Vanishing of the $D$-terms enforces $|\bar{\phi}| = |\phi|$.  Assuming
chaotic initial conditions the fields get trapped in the inflationary
valley of local minima at $|S| > S_c = \eta$ and $\bar{\phi} = \phi =
0$. The potential is dominated by a constant term $ V_0 = \kappa^2
\eta^4 $ which drives inflation.  Inflation ends when the inflaton
drops below its critical value $S_c$ (or when the second slow-roll
parameter $\eta$ equals unity, whatever happens first) and the fields
roll toward the global SUSY minima of the potential $|\phi| =
|\bar{\phi}| = \eta$ and $S=0$.  During this phase transition $B$-$L$
cosmic strings form \cite{prd,lept}. For a discussion on various GUT
models see Ref. \cite{B-L}.

The scalar potential in Eq.~(\ref{Vsusy}) gets corrections from SUSY
breaking by the finite energy density in the universe during inflation
(given by the Coleman-Weinberg formula), from SUSY breaking today, and
from supergravity.  The hidden sector expectation values responsible
for low energy SUSY breaking can generically be written as $\langle z
\rangle = a \mpl, \quad \langle W_{\rm hid} \rangle = \mu \mpl^2,
\quad \langle \frac{\partial W_{\rm hid}}{\partial z} \rangle = c \mu
\mpl$, with $z$ a hidden sector field, $a,c$ dimensionless numbers,
and $\mu$ a mass parameter related to the gravitino mass via $m_{3/2}
= \e^{|a|^2/2} \mu$.  Further $\mpl = (8\pi G)^{-1/2} = 2.4 \times
10^{18} \GeV$ is the reduced Planck mass. The scalar potential along
the inflationary valley can be calculated using the SUGRA formula
\be
V = \e^{K/\mpl^2} \left[ \sum_\alpha
\Big| \frac{\partial W}{\partial \phi_\alpha} 
+ \frac{\phi_\alpha^* W}{\mpl^2} \Big|^2
- 3 \frac{|W|^2}{\mpl^2}
\right].
\label{Vsugra}
\ee
Assuming a minimal K\"ahler potential, the scalar potential including
all corrections is \cite{CMB}
\bea V &=& \kappa^2 \eta^4  \nonumber \\
&+& \frac{\kappa^4 \eta^4 \mcN}{32 \pi^2}
\Big[ 2 \ln\(\frac{2 \kappa^2 \sigma^2}{\Lambda^2}\) + (z+1)^2
\ln(1+z^{-1}) + (z-1)^2 \ln (1-z^{-1}) \Big] 
\nonumber \\
&+& \kappa^2 \eta^4 \Big[
{\sigma^4\over 8 \mpl^4} + {|a|^2 \sigma^2\over 2 \mpl^2} \Big]
+ \kappa A m_{3/2} \eta^2 \sigma,
\label{V}
\eea
with $\sigma = |S|/\sqrt{2}$ the normalized real field, $A= 2 \sqrt{2}
\cos(\arg \mu - \arg S)$, and we assume that $\arg S$ is constant
during inflation. The cosmological constant today vanishes for $|c +
a^*|^2 = 3$, and we have dropped subdominant terms. Further we used
the notation $z= x^2 = {|S|^2}/\eta^2 = \sigma^2 /(2\eta^2)$ so that
$z=x=1$ when $\sigma = \sigma_c$.  The first line is the tree level
potential term, the second line is the one-loop Coleman-Weinberg
correction due to SUSY breaking during inflation \cite{Dvasha}, and
the third line are the SUGRA corrections.

The Coleman-Weinberg potential and the non-renormalisable terms are
always present, independent of low energy SUSY breaking.  The $A$- and
mass terms can be made small for a small gravitino mass (as in gauge
mediation), or by tuning $A$ and/or $a$.  Another possibility is that
the hidden sector superfield $z$ only acquires its VEV after
inflation, so that these terms are absent during inflation. We note
that, assuming gravity mediated SUSY breaking, the generic values
$m_{3/2} \sim 10^2 \GeV$, $A\sim 1$, $a\sim 1$, give rise to too large
$A$- and mass terms, incompatible with the CMB data~\cite{CMB}.

Both strings and the inflaton contribute to the primordial density
perturbations \cite{prd,CMB}. Cosmic strings do not predict the
measured acoustic peaks in the CMB and hence their contribution to the
temperature fluctuations should be small, less than about 10\%
~\cite{Pogosian}.  The string contribution is proportional to the
string tension
\be 
\mu = 2 \pi \eta^2 \theta(\beta),
\label{mu}
\ee
with $\beta = (m_\phi / m_A)^2$.  The Higgs mass is $m_\phi^2 =
\kappa^2 \eta^2$ and the vector boson mass is $m_A^2 \simeq g_{\rm
GUT}^2 \eta^2$ with the GUT coupling $g_{\rm GUT}^2 \approx 4\pi/25$.
When $\beta =1$, the strings satisfy the Bogomolny bound (this is the
case of cosmic strings which form at the end of brane inflation) and
$\theta(1) = 1$. However in the case of SUSY GUTs, the strings never
satisfy the Bogomolny bound, $\beta < 1$ always, and ~\cite{Hill}
\be
\theta(\beta) \approx \left \{
\begin{array}{lll}
1.04  \beta^{0.195}, & \qquad \beta > 10^{-2}, \\
{\displaystyle \frac{2.4}{ \log(2/\beta)}}, & \qquad \beta < 10^{-2}.
\end{array}
\right.
\label{theta}
\ee
Requiring the string contribution to the quadrupole to be less than
10\% gives the bound \cite{CMB}
\be
G \mu < 6.9 \times 10^{-7} \( \frac{3}{y} \) 
\quad \Rightarrow \quad
\eta_{\rm bnd} < 4.1 \times 10^{15} 
\sqrt{\frac{(3/y)}{\theta(\beta)}}.
\label{Pogosian}
\ee
Here $y$ parameterizes the density of the string network, and should
be taken form numerical simulations. Recent work predicts $y=9 \pm
2.5$ \cite{Landriau}.  Older simulations give $y = 6$~\cite{Allen},
and semi-analytic approximations give $y = 3 -6$~\cite{approx}.

\begin{figure}
\begin{center}
\leavevmode\epsfysize=9cm \epsfbox{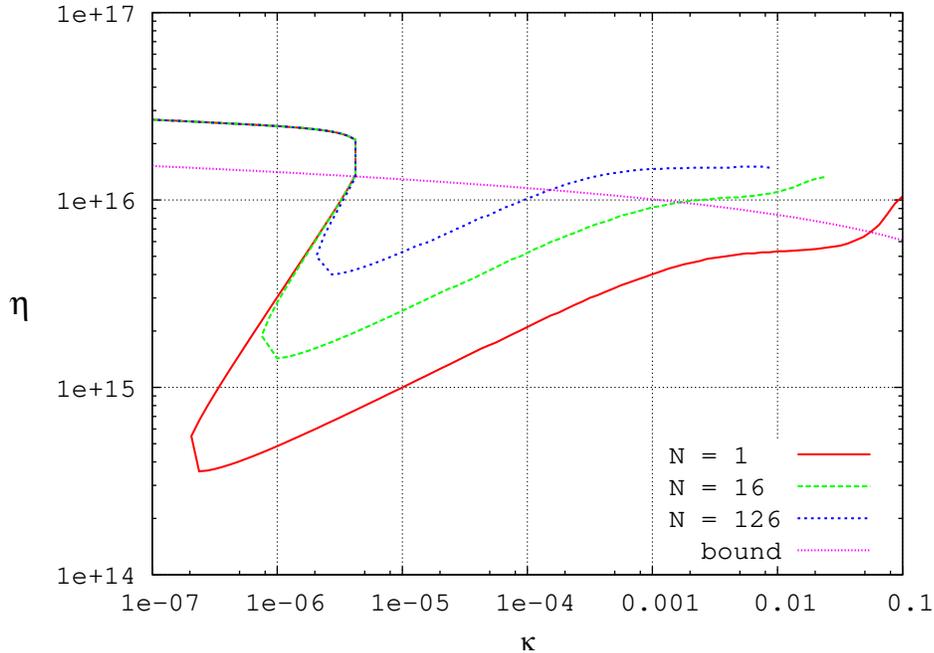}
\caption{$\eta$ vs. $\kappa$ for $\mcN =1,16,126$ and $y=3$. Further
shown is the 10\%-bound.}
\label{F:y3}
\end{center}
\end{figure}

The density perturbations produced in hybrid inflation can be
calculated using the slow roll formalism for the potential in
Eq.~(\ref{V}). Setting it equal to the value observed by WMAP gives
$\eta$ as a function of $\kappa$.  We will use the analytic
approximations, derived in the limit that the
Coleman-Weinberg(CW)-potential respectively the non-renormalisable
(NR) terms dominate the potential:~\cite{CMB}
\bea
\eta_{_{\rm CW}} 
&=& 5 \times 10^{15} \GeV\, \mcN^{1/3}
\( \frac{\kappa}{10^{-3}} \)^{1/3} ,
\label{eta_l}
\\
\eta_{\rm NR} &=& 
3 \times 10^{15} \GeV\, \( \frac{\kappa}{10^{-6}} \) .
\label{eta_NR}
\eea
The symmetry breaking scale is restricted to the range
\be
\eta_{_{\rm CW}} \leq \eta \leq \min[\eta_{\rm NR},\eta_{\rm bnd}].
\label{eta}
\ee
If the $A$- and mass terms are absent or subdominant during inflation
there are two distinct solutions, corresponding to $\eta_{_{\rm CW}}$
and $\eta_{\rm NR}$ (the upper bound $\eta_{\rm bnd}$ comes from the
fact that the string contribution to the CMB is limited). If on the
other hand these terms do play a role, the whole range is possible.
This is illustrated in Fig.~\ref{F:y3}, which shows $\eta$ as a
function of $\kappa$ for $\mcN=1,16,126$.  In this plot the $A$- and
mass term are assumed to be negligible; if they are not the solution
is somewhere in the range given by Eq.~(\ref{eta}). The straight part
at relatively large coupling is well approximated by $\eta_{_{\rm
CW}}$. At low value there is a second branch of solutions given by
$\eta_{\rm NR}$.  Also plotted is $\eta_{\rm bnd}$ for $y=3$; above
this line the string contribution to the CMB is more than 10\%. The
CMB constraints are satisfied for the coupling range $10^{-6} \lesssim
\kappa \lesssim 10^{-2}/\mcN$ and the SSB scale range $\eta \sim
10^{15}-10^{16} \GeV$.

The CMB bound can be avoided if the strings are semi-local or not
topologically stable down to low energy  and decay at some
later phase transition \cite{B-L}.

\section{Non-thermal leptogenesis from inflaton decay}
\label{s:NT_inflaton}

In this section we review the non-thermal leptogenesis scenario which
happens during reheating as a result of inflaton decay into
right-handed (s)neutrinos \cite{leptinfl,shafi1}.

The Higgs fields $\phi$ and $\bar{\phi}$ break local $B$-$L$
spontaneously when developing a VEV at the end of inflation. The
right-handed neutrinos acquire super-heavy Majorana masses via their
coupling to $\bar{\phi}$. Some other GUT superfield (this is model
dependent) gives a Dirac mass to the neutrinos and the light neutrinos
acquire a super light Majorana mass via the see-saw mechanism
\cite{seesaw}. The lepton asymmetry, generated as the right-handed
(s)neutrinos decay into SM Higgs(inos) and (s)leptons, is converted
into a baryon asymmetry via sphaleron transitions \cite{Yanagida}.

Depending on the transformation properties of the Higgs
representation, the right-handed neutrino masses are generated via
normalisable or non-renormalisable superpotential terms
\bea
W &=& \frac{1}{\mpl} \gamma_{ij} \bar{\phi} \bar{\phi} F_i F_j,
\label{Mi_NR} \\
W &=& y_{ij} \bar{\phi} F_i F_j,
\label{Mi_ren}
\eea
where $F$ is the $n$-dimensional spinorial representation of G which
contains the right-handed neutrino superfield $N$, and $i,j = 1..3$
for three families. At the end of inflation $S$ and $\phi_+ = (\delta
\phi + \delta \bar{\phi})/\sqrt{2}$, with $\phi =\eta + \delta \phi$
and $\bar{\phi} =\eta + \delta \bar{\phi}$, oscillate around the
global SUSY minimum of the potential until they decay into
right-handed neutrinos and sneutrinos, thereby reheating the universe
\cite{leptinfl}.  We work in the basis where the right-handed
neutrinos mass matrix is diagonal. The decay rates $\Gamma(\phi_+ \to
N_i N_i)$ and $\Gamma(S \to \tilde{N_i} \tilde{N_i})$ are equal and
given by
\be
\Gamma_N = \frac{1}{8 \pi} \(\frac{M_i}{\eta} \)^2 m_\chi,
\label{gamma}
\ee
with $\chi = S,\phi_+$ the oscillating fields which have equal mass
$m_\chi = \kappa \eta$, and $M_i = y_i \eta, \gamma_i
\frac{\eta^2}{\mpl} $ the mass of the heaviest right-handed neutrino
$N_i$ (sneutrino $\tilde{N_i}$) the inflaton can decay into (i.e.,
which satisfies $M_i < m_\chi/2$).  The reheat temperature is then
\be
T_{\rm R} = \({45\over 2 \pi^2 g^*}\)^{1/4} (\Gamma_N \, \mpl)^{1/2} 
\simeq 6\times 10^{-2} M_i \sqrt{ \frac{\kappa \mpl}{\eta}} .
\label{TR}
\ee
where we have used $g_* = 228.75$ for the MSSM spectrum.

Non-thermal leptogenesis from inflaton decay takes place if the
following constraints are satisfied:

\begin{description}
\item [Kinematic constraint:]

Inflaton decay into right-handed (s)neutrinos is kinematically allowed
if $M_i \leq m_\chi /2$, i.e.,
\be
M_i (\kappa) \leq \frac12 \kappa \eta
\ee

\item [Gravitino constraint:]

Gravitino overproduction is avoided if the reheat temperature $T_{\rm
R} \lesssim 10^{10} \GeV$ \cite{gravitino}. This gives
\be
M_i (\kappa) \lesssim 1.6 \times 10^{11} \GeV \( \frac{T_{\rm R}}{10^{10} \GeV}\)
\sqrt{\frac{\eta(\kappa)}{\kappa \mpl}}.
\label{gravitino}
\ee
The upper bound on $T_{\rm R}$ is model dependent and can be as low as
$10^6 \GeV$.  We note that the gravitino constraint can be avoided if
the gravitino mass is sufficiently large so that it decays before
BBN.

\item [Gravitational decay:]

The decay rate into right-handed (s)neutrinos should be larger than
the gravitational decay rate into light particles. In a full theory,
the superpotential is $W= W_{\rm infl} + W_{\rm hid} + W_{\rm GUT}$,
where $W_{\rm GUT}$ contains GUT superfields, some of them containing
the MSSM fields.  The gravitational decay rate of the inflaton into
light SM particles can then be computed by considering for example a
term of the form $W_{\rm GUT} = a H F F$, $F$ containing the standard
model fermions and $H$ some GUT Higgs superfield containing the SM
Higgs. In the SUGRA potential Eq.~(\ref{Vsugra}) there is a coupling
between the inflaton and the SM particles, leading to a decay rate
\be
\Gamma_{\rm grav} \simeq \frac{1}{8\pi} \frac{ m_\chi^3 \eta^2}{\mpl^4}. 
\label{grav}
\ee
Note that this is parametrically smaller than the standard
gravitational decay rate $\Gamma_{\rm grav} = (1/8\pi)
m_\chi^3/\mpl^2$ \cite{Nanopoulos}. Requiring $\Gamma_N
>\Gamma_{\rm grav}$ then leads to
\be
M_i(\kappa) \gtrsim \frac{\kappa^2 \eta^3}{\mpl^2}.
\label{grav_decay}
\ee

\item [The wash-out constraint:]

The lepton asymmetry produced by the decay of the RH neutrino $N_i$ is
washed out by the $L$-violating processes involving RH neutrinos,
unless they are out of thermal equilibrium which is automatic if $M_j
\lesssim T_{\rm R}$, with $j=1,2,3$. The strongest constraint is for
the lightest RH neutrino:
\be \frac{M_1(\kappa)}{M_i(\kappa)} \gtrsim \frac{1}{16} \sqrt{\frac{\mpl
\kappa}{\eta(\kappa)}},
\label{wash}
\ee
with as before $M_i$ the mass of the RH neutrino the inflaton decays
into.  This implies $M_1 \gtrsim 10^{-1}-10^{-3} M_i$ for ($\eta,
\kappa$) allowed by CMB data (see Eqs.~\ref{eta_l}-\ref{eta}). No
wash-out is assured if $M_1 < m_\chi < M_2,M_3$ and the inflaton
decays into the lightest right-handed neutrino.  The CP violating
parameter $\epsilon$ can be improved by at most a factor $(M_2/M_1)
\simeq 10^1-10^3$ if the decay is not into the lightest neutrino, but
into the next to lightest one (see the Appendix: $\epsilon_1 \propto
M_1$ and $\epsilon_2 \propto M_2$)~\cite{shafi1}.

\item [Perturbative couplings:]

We require the couplings $\gamma_i, y_i$ in
Eqs.~(\ref{Mi_NR},~\ref{Mi_ren}) to be less than unity. For a
renormalisable mass term this bound cannot compete with the kinematic
constraint.  For a non-renormalisable mass term this implies
\be
M_i \lesssim
\frac{\eta^2}{\mpl} \sim 1 \times 10^{15} \GeV 
(\mcN \kappa)^{2/3} ,
\label{pert}
\ee
where in the last step we used $\eta = \eta_{_{\rm CW}}$, see
Eq.~(\ref{eta_l}). If $\eta/\mpl < \kappa$, which only happens for $k
\gtrsim 10^{-2}$, then for non-renormalisable mass terms all neutrino
masses are lighter than the inflaton mass.

\item [Lepton asymmetry:]

The lepton asymmetry produced is~\cite{Asaka}
\be
\frac{n_L}{s} = \frac32 \frac{T_{\rm R}}{m_\chi} \epsilon_i,
\ee
with $\epsilon_i$ the CP asymmetry per decaying RH neutrino $N_i$.
For hierarchical RH neutrino masses and hierarchical light neutrinos
the CP asymmetry in the decay of the lightest RH neutrino is bounded
by \cite{Davidson,Buchmuller}
\be
|\epsilon_1| \leq 2 \times 10^{-10}  \( \frac{M_1}{10^6 \GeV} \) 
\( \frac{(\Delta m^2_{\rm atm})^{1/2}}{0.05 \eV}\).
\label{eps}
\ee
As discussed in the appendix, the upper bound on $\epsilon_2$ is of
the same order of magnitude, but with $M_1 \to M_2$ in the above
formula. The CP-asymmetry induced by the decay of the heaviest RH
neutrino is suppressed by a factor $M_2/M_3$.  For quasi-degenerate
light neutrinos ($m_1 \approx m_2 \approx m_3 \gg (\Delta m^2_{\rm
atm})^{1/2}$) the asymmetry is smaller ~\cite{Buchmuller}
\be
|\epsilon_1| \leq  2 \times 10^{-10}  \( \frac{M_i}{10^6 \GeV} \) 
\(\frac{(\Delta m^2_{\rm atm})^{1/2}}{0.05 \eV}\) 
\(\frac{(\Delta m^2_{\rm atm})^{1/2}}{\bar{m}}\),
\ee
with $\bar{m} = 1/3 \sqrt{m_1^2+m_2^2+m_3^2}$.  

If the RH neutrinos are quasi-degenerate and $M_i - M_j \sim \Gamma_i$
the CP-asymmetry is enhanced. The only constraint is then
$|\epsilon_i| < 1$.  Upper bounds on the CP-asymmetry in type II
see-saw models have also been derived for non-degenerate neutrinos,
and are of the same magnitude as Eq.~(\ref{eps})~\cite{king}.

The baryon asymmetry inferred from BBN translates into a primordial
lepton asymmetry given by $n_L/s = 2.4 \times 10^{-10}$ for the MSSM
spectrum. For hierarchical light neutrinos, this is obtained for
\be
M_i(k) \gtrsim {5.4 \times 10^3}
\( \frac{\kappa \eta(\kappa)^3}{\mpl} \)^{1/4} \sqrt{\GeV}.
\ee
For smaller $M_i$ the produced asymmetry is too small.

\end{description}

\begin{figure}[t]
\leavevmode\epsfysize=9cm \epsfbox{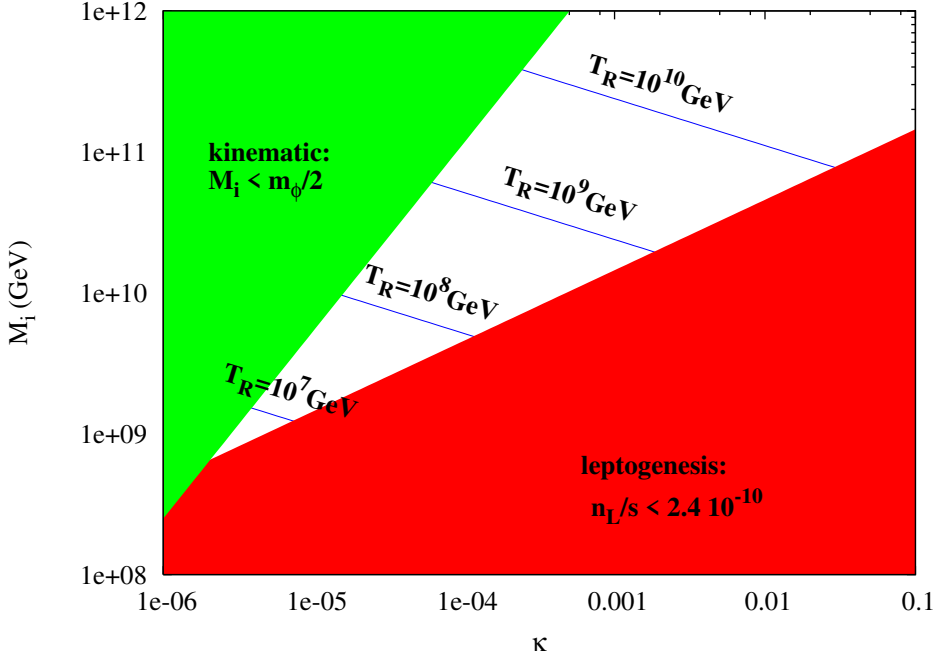}
\caption{$M_i$ vs. $\kappa$ for $\eta= \eta_{_{\rm CW}}$ and
$\mcN=1$. The parameter space is bounded by the kinematic constraint
(left), successful leptogenesis (right), and the gravitino constraint
(lines, for $T_{\rm R} = 10^7,10^8,10^9,10^{10}\GeV$) .}
\label{F:decay1}
\end{figure}

\begin{figure}[t]
\leavevmode\epsfysize=9cm \epsfbox{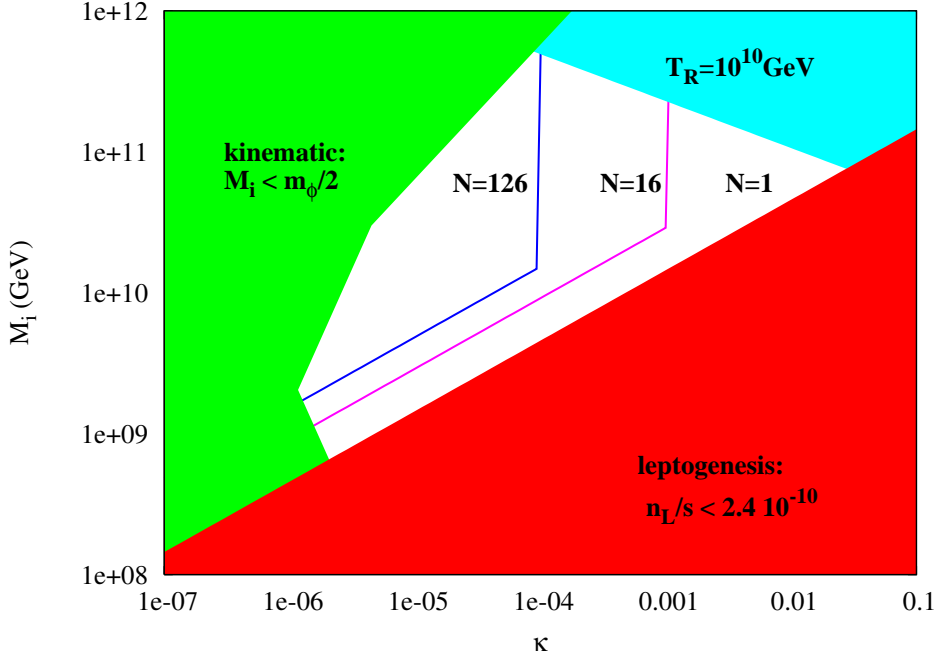}
\caption{$M_i$ vs. $\kappa$ for $\max[\eta_{_{\rm CW}},\eta_{\rm NR}]<
\eta < \eta_{\rm bnd}$, $T_{\rm R} = 10^{10} \GeV$ and $\mcN =
1,16,126$. The parameter space is bounded by the kinematic constraint
(left), successful leptogenesis (bottom), the gravitino constraint
(top), and for $\mcN =16,\,126$ by the CMB data (right).}
\label{F:decay2}
\end{figure}

The parameter space compatible with NT leptogenesis from inflaton
decay is shown in Figs.~\ref{F:decay1} and \ref{F:decay2}.  The
kinematic, perturbative coupling and gravitino constraints all give an
upper bound on $M_i$. The perturbative coupling constraint is weakest
and not shown.  The kinematic constraint is strongest for small
$\kappa$, whereas the gravitino constraint dominates for large
coupling.  The gravitational decay and leptogenesis constraint give a
lower bound on $M_i$.  The leptogenesis bound is strongest.  The
$\kappa$ range is bounded by the CMB data, as given by
Eqs.~(\ref{eta_l}, \ref{eta_NR}, \ref{eta}) and Fig.~\ref{F:y3}, to
$10^{-6} \lesssim \kappa \lesssim 10^{-2}/\mcN$.

Fig.~\ref{F:decay1} shows the parameter space for $\eta= \eta_{_{\rm
CW}}$ and $\mcN = 1$. The colored regions are excluded. The upper
bound is fixed by the reheat temperature; the bounds for $T_{\rm R} =
10^7,10^8,10^9,10^{10} \GeV$ are shown.  Leptogenesis is only possible
for $T_{\rm R} > 10^6 \GeV$.  For $T_{\rm R} \sim 10^9 \GeV$,
successful leptogenesis requires $M_i = 10^9 - 10^{11}\GeV$ and $\kappa
= 10^{-5} - 10^{-3}$.  One should remember that in addition to the
constraints shown in the plot, it should be checked that $M_1 > T_{\rm
R}$ and there is no wash-out of asymmetry.  This is the case for $M_1
\gtrsim 10^{-2} M_i$, with leptogenesis dominated by inflaton decay
into $N_i$, in agreement with Eq.~(\ref{wash}).

For a fixed coupling value, the possible range of neutrino mass $M_i$
is about a decade.  Therefore the bounds are all close to saturation.
E.g. the leptogenesis constraint gives $M_i \propto
(\epsilon_i/\epsilon_i^{\rm max})^{-1/2}$, and thus
$(\epsilon_i/\epsilon_i^{\rm max}) \gtrsim 10^{-2}$ is
required. Hence, degenerate light neutrinos with $\bar{m} \sim \eV$
are marginally excluded.

Fig.~\ref{F:decay2} shows the parameter space for $\eta$ in the whole
range of Eq.~(\ref{eta}) for $T_{\rm R} = 10^{10} \GeV$ and
$\mcN=1,16,126$.  The parameter space is enhanced compared to
Fig.~\ref{F:decay1}, about twice as big.  The main consequence of
increasing $\mcN$ is that the small $\kappa$ range is excluded by the
CMB data, and that leptogenesis requires a slightly larger neutrino
mass.

\section{Leptogenesis from string decay}

Cosmic strings form in SUSY GUT models with standard hybrid inflation
\cite{prd,jrs}. The string mass per unit length is then constrained by
CMB data, see Eqs.(\ref{mu}), (\ref{eta_l}) and (\ref{eta_NR}).  When
the symmetry broken is gauged $B$-$L$, they also provide a non-thermal
scenario for leptogenesis \cite{lept}.  We first describe various
possible NT leptogenesis scenarios with $B$-$L$ strings forming at the
end of inflation. We then discuss the evolution of the string network
and analyse the various scenarios in details.

\subsection{Leptogenesis scenarios}

Depending on whether the mass of the lightest RH neutrino $M_1$ is
larger or smaller than the reheat temperature $T_{\rm R}$, and on
whether the inflaton decays into right handed neutrinos or not, there
will also be a contribution to the lepton asymmetry of the universe
from inflaton decay or standard thermal leptogenesis. We distinguish
the following cases:

\begin{description}
  
\item [Case 1:] The reheat temperature is lower than the lightest
  right-handed neutrino mass, $T_{\rm R} < M_1$, and there is no
  wash-out at any time.  The lepton asymmetry is set by the earliest
  time that string loops form, which is right at the end of inflation.
  Reheating of the universe takes place at a later time, via inflaton
  decay into right-handed neutrinos.  Apart from NT leptogenesis from
  $B$-$L$ strings, there is also a contribution from non-thermal
  leptogenesis from inflaton decay as discussed in section
  \ref{s:NT_inflaton}.

Constraints: gravitino constraint $T_{\rm R} < 10^{10} \GeV$,
gravitational decay constraint $\Gamma_{\rm grav} < \Gamma_N$,
kinematical constraint $M_i < m_\chi/2$.

\item [Case 2:] Same as case 1 but now the inflaton does not decay
  into RH neutrinos. For example, if $M_i > m_\chi/2$ $\forall i$ and
  there is no other superpotential term involving the singlet or the
  Higgs fields, decay is through gravitational interactions: $\Gamma_N
  < \Gamma_{\rm grav}$. Gravitational reheating can alleviate the
  gravitino constraint.  Leptogenesis comes from the decay of the
  string forming gauge field in right handed neutrinos.  This is the
  only contribution to the lepton asymmetry.

Constraint: gravitino $T_{\rm R} < 10^{10} \GeV$.

\item [Case 3:] The reheat temperature is higher than the lightest
  neutrino mass $T_{\rm R} > M_1$.  The asymmetry will be washed out
  at high temperatures by $L$-violating processes mediated by $N_1$,
  and can only be created for $T_{\rm R} < M_1$.  The asymmetry is
  then dominated by the loop formation rate at $T \sim M_1$.  There are two
  contributions to the lepton asymmetry: NT leptogenesis from $B$-$L$
  strings and standard thermal leptogenesis.  This is the case
  considered before \cite{Sahu,Gu}.

Constraints: gravitino constraint $T_{\rm R} < 10^{10} \GeV$, kinematical
constraint $M_1 < m_\chi/2$. 

\end{description}

In order to analyse the various scenarios, we study analytically the
evolution of the string network, both in the scaling regime and at the
initial times.

\subsection{String network and neutrino density}

The evolution of a cosmic string network has been extensively studied
over the years \cite{ShelVil}.  Numerical simulations and analytical
studies agree that the string network reaches a scaling regime, in
which the energy-density carried by the network remains a constant
fraction of the total energy density in the universe. The scaling
solution is an attractor solution, and is independent of the initial
string density. This is one of the reasons that string network at
formation has not been discussed much in the literature.  However, the
lepton asymmetry is dominated by the initial time, and thus depends
sensitively on the initial density.  We first discuss the familiar
scaling regime, before turning to a discussion of the initial string
density.

\subsubsection{The scaling regime}
 
To describe the approach to the scaling regime, we introduce the
characteristic length-scale $L$ which sets the correlation length and
the average distance between long strings~\cite{ShelVil}. The energy
density in long strings, $\rho_\infty \sim \mu/L^2$, evolves as
\be
\dot{\rho}_\infty = -2H\rho_\infty -f(p) \frac{\rho_\infty}{L}
\label{dot_rho}
\ee
where the terms on the right hand side describe the energy loss due to
expansion of universe, and due to production of loops respectively.
The function $f(p)$ depends on the reconnection probability $p$ as
$f(p) \sim \sqrt{p}$. For gauge field theory cosmic strings $p =1$.
Introducing $\gamma(t)$ such that
\be
L = \gamma(t) t,
\label{L}
\ee
one finds that the above equation has a stable attractor solution, the
scaling solution. It does not depend on the initial string density:
\be
L = \gamma_s(t) t \equiv \frac{f(P)}{2(1-\beta)} t
\ee
where we wrote $\beta = H t$.  Since $\gamma_s = \sqrt{p}/(2(1-\beta)) =
{\mathcal O}(1)$ is constant, the long strings scale with the horizon.

The scale $L$ characterizes the network on macroscopic scales, but
does not say anything about what happens on the smallest
scales. Simulations show that the long strings have small-scale
wiggles, whose characteristic length also scales with time
\cite{ShelVil}. These wiggles set the typical loop size, which we
parametrise as
\be
l_{\rm loop} \sim \alpha t
\label{lloop}
\ee
with $\alpha \sim (\Gamma G \mu)^n$, and $\Gamma \sim 50$. The
``standard'' value is $n=1$ giving $\alpha = \alpha_1 \equiv (\Gamma G
\mu)$~\cite{ShelVil}.  More recent simulations suggest $n=3/2 \,(5/2)$
in the radiation (matter) dominated era \cite{vilenkin}.  The loop
formation rate $\dot{n}_{\rm loop}$ is set by the requirement that the
string network keeps its scaling solution. The loops loose energy by
emitting gravitational radiation and contract until the loop radius
becomes of the order of the string width, at which point it decays
emitting $X$-particles (with $X = \phi,A,N$, i.e., the string Higgs or
gauge fields, or RH neutrino zero modes).  For $\alpha \lesssim
\alpha_1$, the loop lifetime is less than a Hubble time and we can
then neglect the red shifting between birth and death.  The injection
rate of right-handed neutrinos during scaling is then simply
\be
\dot{n}_N = x_N \dot{n}_{\rm loop} 
\simeq \frac{x_N}{\gamma^2 \alpha p} t^{-4},
\label{rate}
\ee
where $x_N$ is the number of right-handed (s)neutrinos produced per
decaying loop. 

The minimal number of RH neutrinos released per loop is $x_N = 1$.
However, we expect the loop to decay when its radius becomes of the
order of the string width $m_\phi^{-1}$ with a burst of the strings
Higgs and gauge particles \cite{ShelVil}. These in turn (mostly) decay
into RH neutrinos. The number of neutrinos emitted per loop is then of
the order
\be
x_N \lesssim \frac{E_{\rm loop} \big|_{R \sim m_\phi^{-1}}}{m_X} 
\simeq
\frac{(2\pi)^2 \theta(\beta)}{\kappa} \frac{\eta}{m_X}
\label{f_N}
\ee
with $\theta(\beta) \sim 0.1 -1$ given in Eq.~(\ref{theta}).  If the
loops decay mainly into Higgs fields $m_X = m_\phi$ and $x_N \sim
\kappa^{-2}$, whereas if decay is mainly into gauge bosons $m_X = m_A$
and $x_N \sim \kappa^{-1}$.  If the string width at which the loop
decays is smaller than the the inverse Higgs mass $m_\phi^{-1}$, then
$x_N$ is correspondingly smaller.

The lepton density is obtained by integrating Eq.~(\ref{rate}) with a
red shift factor $(a(t_{\rm in})/a(t))^3 = (t_{\rm in}/t)^{3/2}$ to
account for the expansion of the universe. Here it is assumed that
both $t$ and the initial time $t_{\rm in}$ are in the radiation
dominated era following inflaton decay. This gives (using $t \sim
H^{-1}$)
\be
n_L(H)
\simeq \frac{x_N\epsilon_i}{\gamma_{\rm s}^2 \alpha} 
H_{\rm in}^{3/2} H^{3/2}, 
\label{n_L1_tr}
\ee
independent of the reheating temperature.\\

Some simulations find that loops form on the smallest possible scale
(given by the resolution of the simulation). These results suggest
that scaling is maintained mostly by particle emission rather than via
loop formation and subsequent gravitational decay.  This is called the
VHS scenario, after the authors of Ref.~\cite{VHS}. The emitted
$X$-particles are the RH neutrinos themselves, due to the existence of
zero mode solutions, and the string forming Higgs and gauge fields which
then (mostly) decay into RH neutrinos.  This gives a RH neutrino
injection rate \cite{sigl}
\be
\dot{n}_N \simeq f_{_{X}} \frac{\mu}{m_X \gamma^2} t^{-3}.
\label{rate2}
\ee
with $X = \phi,A,N$, and $f_{_{X}}$ the fraction of the energy in
loops that goes into $X$-particles.  Not all of the loop energy can go
into high-energy particles, as this would give too large a diffuse
$\gamma$-ray back ground, in conflict with the EGRET
data~\cite{egret}.  Combining the EGRET bound with the CMB bound
gives\footnote{A similar injection rate and EGRET constraint apply for
large loops which undergo 'quick death', i.e., loops that decay
through many self-intersections into small loops which in turn decay
emitting heavy X-particles.}~\cite{sigl}
\be 
f_{_{X}}
\lesssim 10^{-5}
\( \frac{7 \times 10^{-7}} {G\mu} \) 
\ee
We note, however, that the EGRET flux is dominated by late times. It
is therefore not impossible that $f_{_{X}}$ is time-dependent,
and much larger than the bound above at early times.

When leptogenesis takes place after reheating, the lepton number
density is obtained by integrating (\ref{rate2}) taking into account
the expansion of the universe.  This gives
\be
n_L(H) 
\simeq  \frac{f_{_{X}} \epsilon_i} 
{\gamma_{\rm s}^2}  \frac{\mu H_{\rm in}^{1/2} H^{3/2}}{m_X} .
\label{n_L2_tr}
\ee

\subsubsection{The initial string density}
\label{sec-init}

Cosmic strings are formed during the $B$-$L$ breaking phase
transition. The string density is set by the correlation length at the
time of the phase transition $\hat{\xi}$~\cite{Kibble,Zurek}.  The
universe is cold at the end of inflation, and the equilibrium
correlation length is set by the mass of the symmetry breaking Higgs
fields $\xi(t)^{-1} =m_\phi(t) = \kappa^2(S^2(t)-\eta^2)$.  Both the
correlation length and the relaxation time $\tau = \xi$ diverge during
the phase transition, and eventually $\phi$ must fall out of
equilibrium.  The correlation length at freeze out $\hat{\xi}$ is thus
in the range
\be
(\kappa \eta)^{-1} < \hat{\xi} < H_*^{-1}.
\label{xi_range}
\ee
with $H_*$ the Hubble constant at the end of inflation.
The lower bound is set by the maximum Higgs mass $m_\phi = k \eta$
obtained in the vacuum.  The upper bound is set by causality, as
fluctuations cannot exceed the horizon.  A more careful estimate of
$\hat{\xi}$ is given in Ref.~\cite{Copeland}.  Writing $S(t) = S_c -
\dot{S} t $ near the phase transition, the inverse Higgs mass, which
sets the correlation length, is $m_\phi^2(t) = - (\kappa^2 \eta
\dot{S}) t$. Freeze out happens approximately at the time when the
relaxation time is equal to $|t|$, and thus
\be
\hat{\xi} \approx \xi(-\tau) = (\kappa^2 \eta \dot{S})^{-1/3}.
\label{xi_best}
\ee
During slow roll inflation $\dot{S} \sim (60 H^2)$.  If the velocity
does not change much between time observable scales leave the horizon
and the time of the phase transition, then this is a good estimate.
The freeze-out correlation length $\hat{\xi}$ determines the typical
distance between cosmic strings.

At the time of string formation (quantities will be denoted by subscript
$_*$), the correlation length $\hat {\xi} = L = \gamma_* t_*$ (see
Eq.~(\ref{L}))
\be
t_*^{-1} \sim H_* \sim \frac{\kappa \eta^2}{\mpl}.
\label{Hstar}
\ee
Using Eqs.~(\ref{xi_range},~\ref{xi_best}), we get the range and
``best'' value for $\gamma_*$:
\be
\gamma_{\rm min} \equiv \eta/\mpl < 
\gamma_* < 1 \equiv \gamma_{\rm max},
\qquad\qquad
\gamma_{\rm best} \equiv \gamma_* \sim 0.1
\kappa^{-1/3} (\eta/\mpl)^{1/3}.
\label{gam_best}
\ee

Since $\gamma_* < \gamma_{s}$ and $\rho_\infty \propto \gamma^{-2}$,
the energy density in the network is initially larger than during the
scaling regime.  Solving Eq.~(\ref{dot_rho}) with the initial
conditions above, it can be seen that the scaling regime is typically
reached in only a couple of Hubble times.  We will use the
approximation that initial network reaches the scaling regime
instantaneously, and an amount of energy $\rho_* \sim \mu/(\gamma_*
t_*)^2$ is dumped into loops and/or particles at the initial time
given by Eq.~(\ref{Hstar}).~\footnote{We neglect friction in the
thermal bath which is absent for $T < G\mu \mpl$. Note that the
strings form before reheating has completed. Taking for $T$ the reheat
temperature Eq.~(\ref{TR}), this gives $M_i < 4\theta
\eta^{5/2}/(\kappa^{1/2} \mpl^{3/2}) \sim 6 \times 10^{14} \GeV \theta
\kappa^{1/3} \mcN^{5/6}$, where in the last step we used
$\eta=\eta_{_{\rm CW}}$.}

Eq.~(\ref{lloop}) for the loop size breaks down at the initial time.
The reason is that the loop radius is smaller than $\sim m_\phi^{-1}$,
($m_A^{-1}$), i.e., the width of the profile function of the Higgs
(gauge) field, and various parts of the loop overlap.  It is not
possible to speak of cosmic string loops anymore, which are well
defined only for $l_{\rm loop} \sim \alpha t > 2 \pi m_\phi^{-1},
(2\pi m_A^{-1})$, which requires $t \gg t_*$. Therefore it is expected
that initially the energy loss in gravitational radiation is small,
and the network mainly decays directly into RH neutrinos and into
gauge and Higgs fields which (mostly) decay into RH neutrinos. The
lepton number density at $H<T_R$ is then
\be n_L(H) \simeq f_{_{X}} \epsilon_i \frac{\rho_*}{m_X} \(
\frac{a(t_*)}{a(t)} \)^3 \simeq  \frac{f_{_{X}} \epsilon_i}
{\gamma_*^2} \frac{\mu \Gamma^{1/2} H^{3/2}}{m_X}.
\label{n_L2}
\ee
where $f_{_{X}}$ is the fraction of the energy going into X-particles and
$\epsilon_i$ is the CP asymmetry per decaying RH neutrino, see
Eq.~(\ref{eps}).  Before inflaton decay the universe is matter
dominated and $a \propto t^{2/3}$, afterwards the universe is
radiation dominated and $a \propto t^{1/2}$.  This is used in the
second step to write $(a(t_*)/a(t))^3 = H^{3/2} H_*^{-2}
\Gamma^{1/2}$, where the time $t \sim H^{-1}$ is after reheating of
the universe $H < \Gamma$.  This factor takes the red shift due to the
expansion of the universe into account.  The earlier the transition
from matter to radiation domination, i.e., the larger the decay rate,
the larger $T_R$ and the larger is the final number density.  The
initial time $t_{\rm in}$ is right at the end of inflation, see
Eq.~(\ref{Hstar}).

Eq.~(\ref{n_L2}) is the same as the lepton asymmetry produced during
the scaling regime as given by Eq.~(\ref{n_L2_tr}) under the
replacement $\gamma_* \to \gamma_s$ (difference in energy densities
stored in long strings), and $\Gamma \to H_{\rm in}$ (difference in
whether leptogenesis takes place before or after reheating).

\subsection{Results}

For the parameters at hand $\Gamma < H_*$ and the inflaton decays some
time after inflation.  The lepton asymmetry ${n_L}/{s}$ is to be
evaluated after reheating, when entropy is defined.  The entropy is
$s=(2\pi^2/45)g_*T^3 \sim 10^2 T^3$ with $T \simeq 0.4\sqrt{\mpl H}$.

\subsubsection{Case 1}

The string contribution is dominated by the asymmetry produced at the
initial time.  The lepton number density is given by Eq.~(\ref{n_L2}).
Dividing by the entropy gives
\be
\frac{n_L}{s} \simeq \frac{\epsilon_i  f_{_{\rm X}}}{\gamma_{\rm in}^2}
\frac{\mu \Gamma_N^{1/2}}{m_X \mpl^{3/2}}.
\label{eta1}
\ee
Using Eqs. (\ref{gamma}) and (\ref{eps}), we see that $n_L/s$ is
proportional to $M_i M_j$ the mass of the heaviest RH neutrino the
inflaton can decay into and the mass of the RH neutrino which is
mostly produced by strings.  For $M_i=M_j$, which is automatic if the
strings decay mostly into $\phi$-particles, successful leptogenesis
with $n_L/s =2.4 \times 10^{-10}$ requires
\be
M_i \simeq 9 \times 10^3 C
\( \frac{\mpl^{3} \GeV^{2}}{\eta \kappa} \)^{1/4}
= \frac{4 \times 10^{12} C \GeV}{\kappa^{1/3} \mcN^{1/12}} ,
\label{M1}
\ee
where in the second step we used $\eta=\eta_{_{\rm CW}}$, and we
introduced
\be 
C = \gamma_*
\sqrt{ \frac{m_X/\eta}{\theta(\beta) f_{_{X}}}}.
\label{C}
\ee
$C$ is minimized for $\gamma_* \to \gamma_{\rm min}$ and $f_{_{X}} \to
1$.  In this limit the energy density in the string network is of the
same order as the energy density in the oscillating inflaton field.
If both the inflaton and the strings decay into $\phi$ particles ($X =
\phi$) this gives a contribution to the lepton asymmetry of similar
magnitude.  If however $\gamma_* > \gamma_{\rm min}$ or $f_{_{X}} <
1$, the energy stored in the string network is subdominant, and thus
also its contribution to the asymmetry:
\be
\frac{(n_L)_{\rm inflaton}}{(n_L)_{\rm strings}} \sim 
f_{_{X}} \(\frac{\gamma_*}{\gamma_{\rm min}}\)^2
\ee

The string contribution can dominate over the inflaton contribution to
the asymmetry if the strings decay mostly into RH (s)neutrinos.
Another possibility is that $M_1 < m_\chi < M_2$ and the string decays
mostly into gauge particles.  The inflaton decays into lightest RH
(s)neutrino, whereas the gauge field can also decay in the next to
lightest one.  Since $\epsilon_2/\epsilon_1 \sim M_2/M_1$ the
CP-asymmetry is then larger per decaying $A$-particle than per
decaying $\phi$-particle.  However, the number of gauge particles
produced by string decay can be of the same order as the number of
Higgs particles from inflaton decay only in the limit $\kappa \to 1$
(so that $m_\phi \sim m_A$) and $f_{_{X}} \to 1$.

For degenerate light neutrinos the CP asymmetry $\epsilon_i$ is
smaller by a factor $\Delta m_{\rm atm}/m_3$, and $M_i \propto
\epsilon_i^{1/2}$ is larger.  The $\mcN$-dependence of the neutrino
mass $M_i$ is weak; the main effect of considering a larger Higgs
representation is the stronger constraint on the coupling coming from
CMB data: $\kappa \lesssim 10^{-2}/\mcN$.  The bound on the neutrino
mass is weakest in the limit $\kappa \to 1$.  This is only possible if
the strings do not contribute to the CMB anisotropies, i.e., if they
are semi-local or are not topologically stable \cite{B-L}.

\subsubsection{Case 2}

\begin{figure}[t]
\leavevmode\epsfysize=9cm \epsfbox{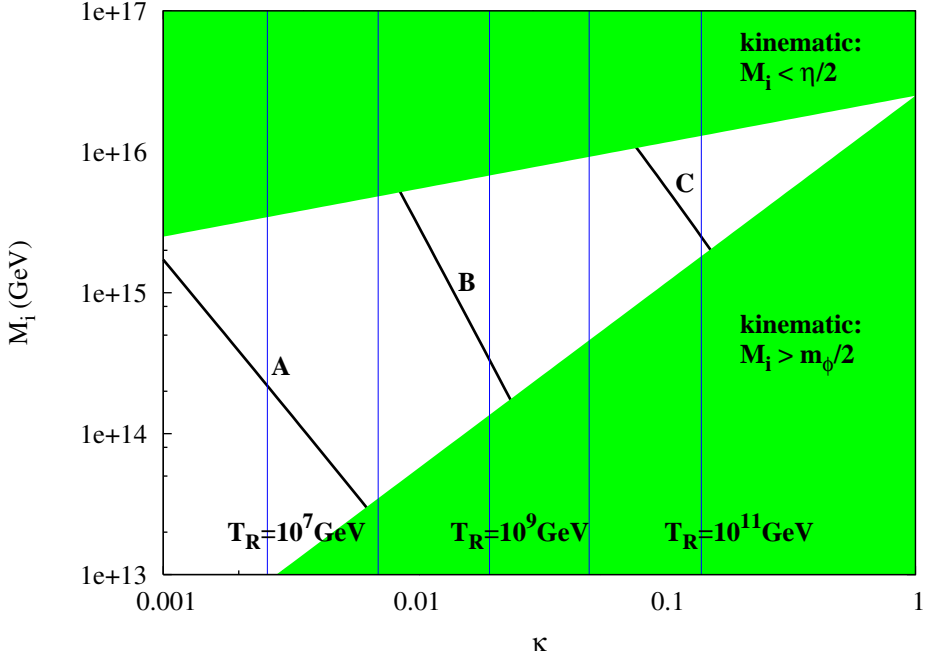}
\caption{$M_i$ vs. $\kappa$ for case 2 with $\eta = \eta_{_{\rm CW}}$
and $\mcN=1$. The lines A-C correspond to $A = (\gamma_{\rm
in},X,f_{_{\rm X}}) = (\gamma_{\rm min},A,1)$, $B = (\gamma_{\rm
best},A,1)$, $C = (\gamma_{\rm min},A,10^{-5})$.  The parameter space
is bounded by the kinematic constraint (top and bottom), and gravitino
constraint (parallel lines, for $T_R =10^7,...,10^{11} \GeV$).}
\label{F:case2}
\end{figure}

If $m_\chi < M_i< \eta$ inflaton decay into RH neutrinos is
kinematically forbidden, and in the absence of other direct couplings
the inflaton decays gravitationally.  This implies a low reheat
temperature which alleviates the gravitino constraint.  Note however
that the lepton asymmetry $n_L \propto \Gamma^{1/2}$ is less efficient
and large neutrino masses are needed.  The string loops decay into
RH neutrinos and gauge quanta which subsequently decay into RH neutrinos.

To get the RH neutrino mass required for leptogenesis in case 2, we
use Eq.~(\ref{eta1}) with $\Gamma_N$ replaced by the gravitational
decay rate, to yield
\be
M_i \simeq 9 \times 10^5 \GeV \frac{C^2}{\kappa^{3/2}}
\( \frac{\mpl}{\eta} \)^{7/2} 
=
\frac{7\times 10^{11} \GeV C^2}{\kappa^{8/3} \mcN^{7/6}},
\label{M2}
\ee
with $C$ given by Eq.~(\ref{C}).  The neutrino mass is quadratic in
$C$, and thus the dependence on the uncertain parameters grouped in
$C$ is larger than in case 1. Minimizing $C$ gives $M_i \gtrsim 2
\times 10^{13} \GeV$.

The lower bound on the neutrino mass Eq.~(\ref{M2}) is shown in
Fig.~(\ref{F:case2}) as well as the kinematic constraint $m_\chi/2 <
M_i< m_A/2 \sim \eta/2$, and the gravitino constraint. The reheat
temperature is independent of the neutrino mass, but does depend on
the coupling $\kappa$.  For $\kappa < 10^{-2}$ one has $T_{\rm R} <
3\times 10^9\GeV $ and there is no gravitino problem. Leptogenesis is
possible for $M_i \sim 10^{14}- 10^{16} \GeV$ and $\kappa \sim
10^{-2}$. The RH neutrino mass increases rapidly with small $\kappa$,
and much smaller couplings are excluded. The large neutrino masses
needed are incompatible with a non-renormalisable mass term as in
Eq.~(\ref{Mi_NR}) and perturbative couplings, see Eq.~(\ref{pert}).

\subsubsection{Case 3}

Consider now the case that the lightest RH neutrino reaches thermal
equilibrium after inflation ($M_1 < T_{\rm R}$).  As follows from
Eq.~(\ref{wash}) thermal equilibrium can only occur if the inflaton
decays in $N_2$ or $N_3$, and not in the lightest RH neutrino.  The
lepton asymmetry is thus produced via string decay in one of the
heavier RH neutrinos and $i=2,3$. Any produced lepton asymmetry will
be washed out until $L$-violating reactions fall out of equilibrium at
$T\sim M_1$.  We assume that this occurs in the scaling regime.

\paragraph{Loop scenario}

The lepton number density is now given by
Eq.~(\ref{n_L1_tr}). Diving by the entropy we get
\be
\frac{n_L}{s} \simeq
\frac{\epsilon_i f_N}{\gamma_{\rm in}^2 \alpha} 
\(\frac{H_{\rm in}}{\mpl}\)^{3/2},
\label{n_L3}
\ee  
with $H_{\rm in} \simeq M_1^2/\mpl$.  Setting it equal to the observed
value gives the RH neutrino mass needed for successful leptogenesis:
\be
M_i \simeq 10^{14} \GeV C^{1/2} \(\frac{M_i}{M_1}\)^{3/4} 
\gtrsim
2\times10^{14} \GeV \(\frac{C'^{2}\mcN^{1/2}}{\kappa}\)^{1/4}.
\label{M3a}
\ee
In the second step we used the equilibrium condition Eq.~(\ref{wash})
and $\eta = \eta_{_{\rm CW}}$; further we defined
\be
C' = \sqrt{ \frac{(\alpha/\alpha_1)}{f_N} } 
\label{C'}
\ee
The number of RH neutrinos released per loop is bounded by $x_N
\lesssim \kappa^{-2}$  (see Eq.~(\ref{f_N})). Further $\alpha$
gives the loop size at birth (see Eq.~(\ref{lloop})); taking $\alpha =
(\Gamma G \mu)^n$ with $n=3/2$ instead of $n=1$ lowers $C'$ by about a
factor 10. Thus $C' \gtrsim 0.1 \kappa$.  The lower bound on the RH
neutrino mass is then
\be
M_i \gtrsim 6 \times 10^{12} \GeV 
\mcN^{1/8} \(\frac{\kappa}{10^{-2}}\)^{1/4},
\ee
together with the CMB constraint $10^{-6} \lesssim \kappa \lesssim
10^{-2} / \mcN$. 

\begin{figure}[t]
\leavevmode\epsfysize=9cm \epsfbox{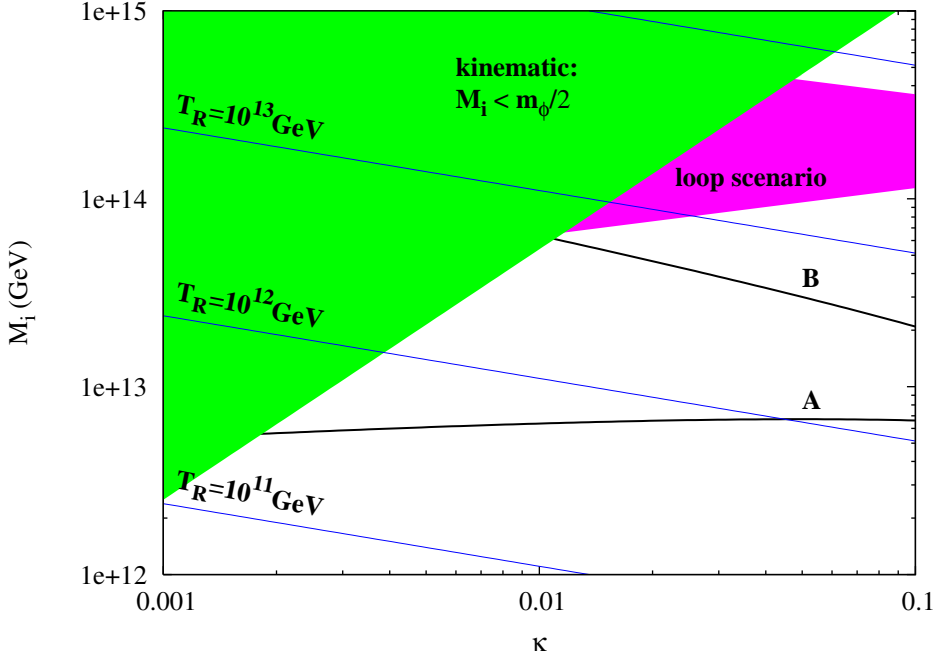}
\caption{$M_i$ vs. $\kappa$ for case 3 with $\eta= \eta_{_{\rm CW}}$,
$\mcN=1$, and $\alpha = \alpha_1$.  The lila part corresponds to
leptogenesis in the loop scenario with $1 < x_N < \kappa^{-2}$, and the
lines A-B to the VHS scenario with $A = (\gamma_{\rm in},X,f_{_{\rm
VHS}}) = (\gamma_{\rm s},\phi,1)$, and $B = (\gamma_{\rm s},A,1)$.
The parameter space is bounded by the kinematic constraint (top), and
the gravitino constraint (parallel lines, for $T_R
=10^{11},...,10^{14} \GeV$).}
\label{F:case3}
\end{figure}

The bound on the neutrino mass Eq.~(\ref{M3a}) is shown in
Fig.~(\ref{F:case3}) for $1 < f_N< \kappa^{-2}$, together with the
kinetic and gravitino constraint.  Large couplings $\kappa \gtrsim
10^{-2}$ are needed, which is marginally excluded by the CMB data. The
reheat temperature has to be large $T_R \gtrsim 10^{13} \GeV$.

\paragraph{VHS Scenario}

The lepton number density is now given by Eq.~(\ref{n_L2_tr}), leading
to
\be
\frac{n_L}{s} = \frac{\epsilon_i  f_{_{X}}}{\gamma_{\rm in}^2}
\frac{\mu H_{\rm in}^{1/2}}{m_X \mpl^{3/2}}
\ee
with $H_{\rm in} \simeq M_1^2/\mpl$, $\gamma_{\rm in} = \gamma_s \sim
1$, and with $i=2,3$.  Leptogenesis requires
\be
M_i \simeq 4\times 10^2 C \frac{\mpl \GeV^{1/2}}{\eta^{1/2}}
\( \frac{M_i}{M_1} \)^{1/2}
\gtrsim \frac{7 \times 10^{12} \GeV C}{\kappa^{1/3} \mcN^{1/12}}
\label{M3b}
\ee
where in the second step we used the equilibrium condition
Eq.~(\ref{wash}) and $\eta = \eta_{_{\rm CW}}$.

The EGRET data requires the fraction of the total energy that goes
into $X$-particles to be small $f_{_{X}} \lesssim 10^{-5}$; as a
result leptogenesis is not efficient enough and too large neutrino
masses are needed, incompatible with the kinematic constraint.  If
however in the early scaling regime $f_{_{X}} \sim 1$ also possible
--- remember that the EGRET bound is determined by late times ---
smaller neutrino masses are possible, as shown in
Fig.~(\ref{F:case3}).

Our results Eq.~(\ref{M3a},~\ref{M3b}) for $C=1,\,C'=1$ agree with
those found in Ref.~\cite{Sahu}.  Ref.~\cite{Gu} assumes degenerate light
neutrino masses, and finds stronger bounds.

\section{$\mu$-term}

The $\mu$-problem can naturally be resolved in SUSY hybrid inflation
with the introduction of a superpotential term \cite{Laz3}
\be W = \lambda S H H'  \ee
where $H,H'$ contains the two Higgs doublets of the MSSM.  After
inflation, $S$ gets a VEV due to low energy SUSY breaking which is of
order $\langle S \rangle \sim m_{3/2}/\kappa$ provided $\lambda >
\kappa$ (otherwise $S$ ends up in wrong minimum), and the $\mu$-term
is generated. However, the superpotential term above opens up a new
decay channel for the inflaton and jeopardizes non-thermal
leptogenesis.

The kinematic constraint $M_i < m_\chi/2$ together with the constraint
$\lambda > \kappa$ assures that decay rate for inflaton decay into SM
higgses and higgsinos
\be
\Gamma_H = \frac{\lambda^2}{16\pi} m_\chi
\label{gammaH}
\ee
is larger than the decay rate into RH neutrinos.  Inflaton decay is
predominantly into SM Higgs fields (unless all the couplings are tuned
$2 M_i/\eta \sim \kappa \sim \lambda$), and NT leptogenesis via
inflaton decay does not occur.  Likewise, string decay into Higgs
fields does not contribute to the lepton asymmetry, since the Higgs
decays into SM Higgses.  On the other hand, if the string decays into
gauge fields, or if RH zero modes are released during decay,
leptogenesis is still possible.  We will consider this possibility in
some detail.  

As a side remark, we note that there are other ways in these models to
generate a $\mu$-term which do not alter the inflaton decay rate, and
are thus compatible with NT leptogenesis from reheating \cite{axion}.

Either case 2 or case 3 is realized, depending on whether the reheat
temperature
\be
T_{\rm R} \simeq 4 \times 10^{-2} \kappa^{3/2} \sqrt{\eta \, \mpl}
\( \frac{\lambda}{\kappa}\)
\ee
is smaller or larger than the mass of the lightest RH neutrino.  Note
that the reheat temperature is minimized in the limit $\lambda \to
\kappa$.

\subsection{Results}
\subsubsection{Case 2}

Leptogenesis is a result of string decay into RH neutrinos and into
gauge fields which in turn decay into RH neutrinos. The RH neutrinos
are out-of-equilibrium at all times, and the asymmetry is dominated by
the initial time just at the end of inflation. There is no
contribution to the asymmetry from inflaton decay nor from thermal
leptogenesis.

\begin{figure}[t]
\leavevmode\epsfysize=9cm \epsfbox{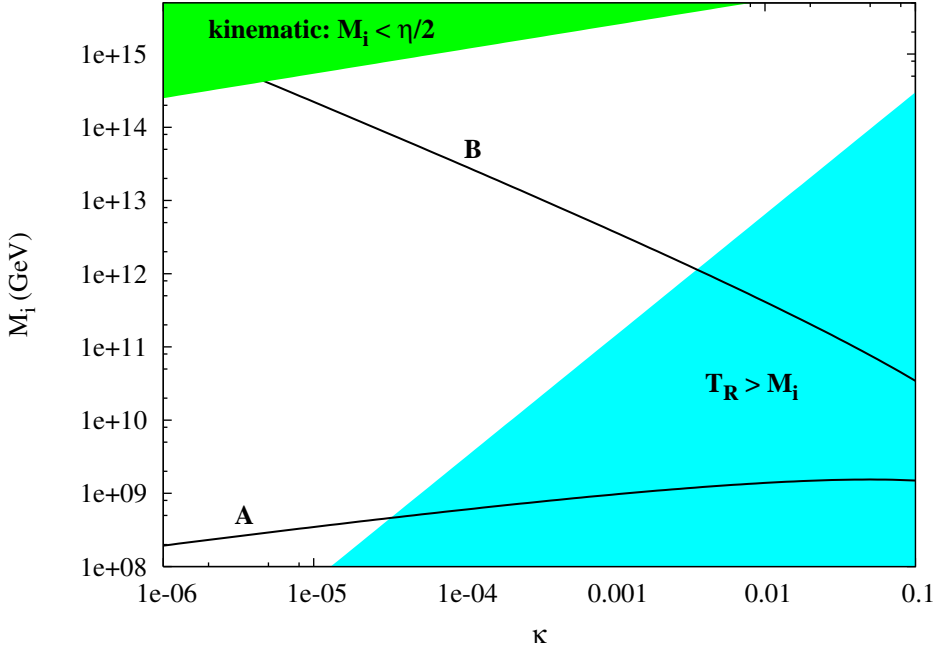}
\caption{$M_i$ vs. $\kappa$ for case 2 with $\mu$-term and $\eta=
\eta_{_{\rm CW}}$, $\mcN=1$.  The lines A-B correspond to $A =
(\gamma_{\rm in},X,f_{_{X}}) = (\gamma_{\rm min},A,1)$, and $B =
(\gamma_{\rm best},A,1)$.  The parameter space is bounded by the
kinematic constraint (top), and the wash-out constraint $T_R < M_1$
(bottom).}
\label{F:nu1}
\end{figure}

The lepton asymmetry is given by Eq.~(\ref{n_L2}) with the replacement
$\Gamma_N \to \Gamma_H$ and $X =A$.  This gives
\be
M_i \simeq \frac{3\times 10^{8} \GeV C^2}{\kappa^{2} \mcN^{1/2}}
\(\frac{\kappa}{\lambda}\)
\ee
where we used $\eta =\eta_{_{\rm CW}}$.  The reheat temperature is a
function of $\kappa$ only: $T_{\rm R} (\kappa/\lambda) =
10^8,10^{10},10^{12} \GeV$ for $\kappa = 10^{-5},2\times 10^{-4}, 3
\times 10^{-3}$. The neutrino mass can be lowered by lowering the
ratio $(\kappa/\lambda) \leq 1$, but at the cost of increasing the
reheat temperature with the inverse ratio. This makes it harder to
satisfy the out-of-equilibrium condition $M_i > T_{\rm R}$.

The results are shown in Fig.~\ref{F:nu1} for various values of $C$
together with the kinematic constraint $M_i < m_A \sim \eta$, and the
out-of-equilibrium condition $M_i > T_{\rm R}$.  Note that the lower
bound on the neutrino mass is proportional to $f_{_{X}}^{-1}$, and the
results for different values of $f_{_{X}}$ can be obtained by
multiplying with the appropriate factor.  Masses as low as $M_i \sim
10^8 \GeV$ are compatible with leptogenesis.

\subsubsection{Case 3}

The lightest RH neutrino reaches thermal equilibrium, and all asymmetry
is erased until it falls out of equilibrium at $T \sim M_1$.

\paragraph{Loop scenario}

The lepton asymmetry is independent of the decay rate, in particular,
on whether the inflaton decays into RH neutrinos or SM Higgses.
Hence, the results of case 3 without a $\mu$-term apply and
\be
M_i \simeq 10^{14}\GeV C'^{1/2} 
\(\frac{M_i}{M_1} \)^{3/4}.
\ee
The only differences are that now $i=1$ is possible and still the lightest RH
neutrino reaches equilibrium, and $f_N < \kappa^{-1}$ so that
$C>\kappa^{1/2}$.  The results do not depend on the decay rate and the
symmetry breaking scale $\eta$; the $\kappa$-dependence enters only
via $x_N$.

\begin{figure}[t]
\leavevmode\epsfysize=9cm \epsfbox{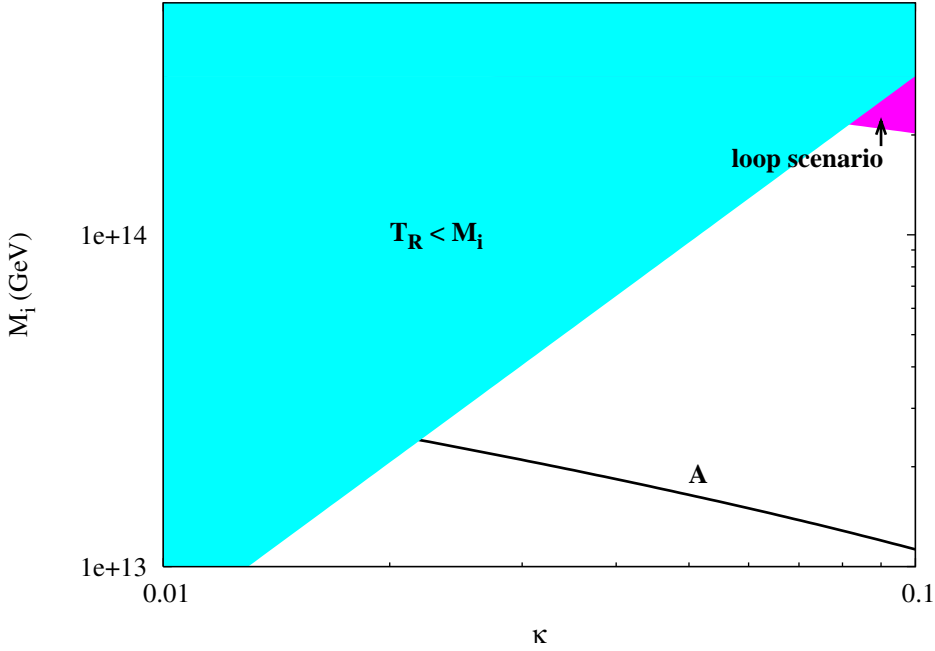}
\caption{$M_i$ vs. $\kappa$ for case 1 with $\mu$-term and $\eta=
\eta_{_{\rm CW}}$, $\mcN=1$.  The lila part corresponds to leptogenesis
in the standard scenario with $1 < f_N < \kappa^{-1}$, and the line A
to the VHS scenario with $(\gamma_{\rm in},X,f_{_{X}}) =
(\gamma_{\rm s},\phi,1)$.  The parameter space is bounded by the
kinematic constraint (top), and the equilibrium constraint $T_R > M_1$
(top).}
\label{F:nu3}
\end{figure}

The lower bound on $M_i$ is shown in Fig.~\ref{F:nu3} for $1 < x_N <
\kappa^{-1}$, together with the kinetic and equilibrium $M_1 < T_R$
constraint. Here it is assumed that $(\kappa/\lambda)=1$, which
minimizes the reheat temperature.  Hence, the equilibrium constraint
can be relaxed by taking $(\kappa/\lambda)<1$, but at the cost of
increasing the reheat temperature, and thereby aggravating the
gravitino problem.

\paragraph{VHS scenario}

The lepton asymmetry is the same as in case 3 without a $\mu$-term,
and requires
\be
M_i \simeq 4 \times 10^{2} \frac{\mpl \GeV^{1/2}}{\eta^{1/2}}
\( \frac{M_i}{M_1} \)^{1/2} =
\frac{4 \times 10^{12} \GeV C' }{(\kappa \mcN)^{1/6}}
\( \frac{M_i}{M_1} \)^{1/2}
\ee
where in the second step we used $\eta = \eta_{_{\rm CW}}$.  The result
is shown in Fig.~\ref{F:nu3}.  The lower bound on the neutrino mass is
proportional $C \propto f_{_{X}}^{-1/2}$, and the result for
different values of $f_{_{X}}$ can be obtained by multiplying
with the appropriate factor.\\

For couplings $\kappa <10^{-4}$ the reheat temperature is $T_R <
10^{10} \GeV$, and this is a scenario with can accommodate inflation
and leptogenesis, and in which both the gravitino problem and the
$\mu$-problem are solved.  The lightest RH neutrino can be in
equilibrium or not: both case 2 and 3 can work.

\section{Conclusions}

In this paper, we have investigated various possibilities for
leptogenesis after hybrid inflation when gauged $B$-$L$ is
spontaneously broken at the end. One of the Higgs fields gives heavy
Majorana mass to the RH neutrinos and NT leptogenesis can take place
during reheating via inflaton decay into RH (s)neutrinos. Cosmic
strings form at the end of inflation~\cite{prd,jrs}.  If stable, they
also contribute to primordial fluctuations. Interestingly enough,
since the string Higgs field breaks $B$-$L$, these are the so-called
$B$-$L$ strings whose decay gives a second NT contribution to the
lepton asymmetry of the universe \cite{lept}. In this paper we
investigated which of these two mechanisms is most efficient, taking
into account the CMB constraints \cite{CMB}.

Leptogenesis via inflaton decay can account for the observed asymmetry
for neutrino masses in the range $M_i = 10^9 - 10^{11} \GeV$, and
quartic Higgs couplings $\kappa = 10^{-5} - 10^{-2}$. The minimal
reheat temperature required for getting enough lepton asymmetry is
$T_R \sim 7 \times 10^6$ GeV.  To assure that all are
out-of-equilibrium at $T_R$, and there is no wash-out of lepton
number, the mass of the RH neutrino the inflaton decays into has to
satisfy $M_i \lesssim 10^2 \times T_R$.

The calculation of the lepton asymmetry created by string decay is
hampered by our poor knowledge of the properties of the string
network.  The initial string density, the loop formation and decay
mechanisms all introduce uncertainties.  In general, only the 'best
case' scenarios give a large contribution to the lepton asymmetry, in
which it is assumed that the initial string density is high and/or
that the fraction of string energy going into $X$-particles is
appreciable.  We argued that both of these assumptions are not
far-fetched, as they fit well with our knowledge of cosmic strings.

We distinguished three different cases, depending on whether the
inflaton field decays into RH (s)neutrinos, and wether the lightest
neutrino is out-of-equilibrium at reheating.

In case 1, the inflaton decays into RH neutrinos and all RH neutrinos
are out-of-equilibrium at $T_R$.  The lepton asymmetry is determined
by the energy density in the string network right at the end of
inflation.  If the strings mostly decay into Higgs particles, the
contribution to the asymmetry is subdominant with respect to the
contribution from inflaton decay.  The strings can give a dominant
contribution if it decays mostly into RH neutrinos or into gauge
particles, but for the latter only in the 'best case' scenario.

If inflaton decay is not into RH neutrinos and the RH neutrinos never
attain thermal equilibrium (case 2), the only contribution to the
baryon asymmetry of the Universe is from cosmic strings decay.  String
decay into RH neutrinos and vector particles, which subsequently decay
into RH neutrinos, can still produce a lepton asymmetry. The minimal
reheat temperature required for getting enough lepton asymmetry $T_R
\sim 10^6$ GeV.  Gravitational inflaton decay can ameliorate the
gravitino constraint.  However, large $M_i \sim 10^{13}-10^{15} \GeV$
RH neutrino masses and couplings $\kappa \gtrsim 10^{-3}$ are
required.  Such large couplings favor small Higgs representations.

The most unfavorable scenario is case 3, in which the lightest RH
neutrino attains thermal equilibrium.  The mass of the heaviest RH
neutrino the inflaton can decay into (which cannot be $M_1$) must be
large, $M_i > 10^{13} \GeV$, the coupling must also be large, and the
reheat temperature must be high $T_{\rm R} \gtrsim 10^{13} \GeV$.  We
note that in this case there is also a contribution to the lepton
asymmetry from thermal leptogenesis.

Finally, we looked at the possibility of generating the MSSM $\mu$-term
dynamically.  Inflaton decay is now into SM Higgses and Higgsinos, and
only case 2 and 3 can occur.  For a wide range of RH neutrino masses
and for couplings $\kappa <10^{-4}$ this is a scenario which can
accommodate inflation and leptogenesis, and in which both the gravitino
problem and the $\mu$-problem are solved.

\section*{Acknowledgements}
                                                                               
RJ would like to thank The Dutch Organisation for Scientific Research
[NWO] for financial support.

\newpage
\appendix
\section{CP-asymmetry}

In this appendix we derive the typical values for the CP-asymmetry
factor $\epsilon_i$, extending the usual analysis to $i=2,3$. We use
the formalism and notation of \cite{smirnov}.

The CP-asymmetry is
\be
\epsilon_i = -\frac{1}{8\pi} \frac{1}{(h h^\dagger)_{ii}}
\sum_{j\neq i} {\rm Im} \[ \{ (hh^\dagger)_{ij} \}^2 \]
 f (\frac{M_j^2}{M_i^2}) 
\ee
with in SUSY theories 
\be 
f(x) = \sqrt{x} \[ \frac{2}{x-1} + \ln(1+x^{-1}) \].  
\ee
For hierarchical RH neutrino masses $M_1 \ll M_2 \ll M_3$, we need the
limits $x \ll 1$ (where $f \to 3/\sqrt{x}$) and $x \gg 1$ (where $f
\to -\sqrt{x}(2+\ln(x))$). The function $f$ is maximized for $f \sim
1$, but this does not occur for hierarchical RH masses.
Then~\cite{smirnov}
\bea
\epsilon_1 &=&
-\frac{3}{8\pi} \frac{1}{(h h^\dagger)_{11}}
\[ 
{\rm Im} \[ \{ (hh^\dagger)_{12} \}^2 \] \frac{M_1}{M_2}
+ {\rm Im} \[ \{ (hh^\dagger)_{13} \}^2 \] \frac{M_1}{M_3}
\] 
\nonumber \\
&\approx& 
-\frac{3}{8\pi} \frac{m_u^2}{v^2}I.
\eea
Here $I \sim {\mathcal O}(1)$ a phase factor, $v = 174 \GeV \sin
\beta$ the Higgs VEV (note that $\sin \beta \approx 1$ for $\beta
\gtrsim 3$).  In the second step we have used
\be
(hh^\dagger)_{22} \approx (m_c^2/ v^2) I_{22},
\quad
(hh^\dagger)_{12} \approx (m_u m_c/ v^2) I_{12},
\quad
(hh^\dagger)_{23} \approx (m_t m_c/ v^2) I_{23},
\ee
with $I$ order one constants, and $m_u, m_c, m_t$ he Dirac neutrino
masses which are labeled in analogy with the quark masses.  One can
express the Dirac masses in terms of the RH neutrino masses:
\be
|M_1| = m_u^2/A_1, \quad |M_2| = m_c^2 /A_2, \quad 
|M_3| = m_t^2/A_3, 
\ee
with $A_1 = s_{12}^2 \sqrt {\Delta m_{sol}^2}$, $A_2 = \sqrt {\Delta
m_{atm}^2}/2$ and $A_3 = 2|m_1|/s_{12}^2$.  The numerical values for
$A_i$ depend on the spectrum of light neutrino masses.  For example
$A_1 =2 \times 10^{-12} \GeV, \, 3 \times 10^{-11} \GeV$ for normal
hierarchy ($m_3 \approx (\Delta m^2_{\rm atm})^{1/2} \gg m_2 \approx
(\Delta m^2_{\rm sol})^{1/2} \gg m_1$) respectively inverted
hierarchy($m_1 \approx m_2 \approx (\Delta m^2_{\rm atm})^{1/2} \gg
m_3 \approx (\Delta m^2_{\rm sol})^{1/2}$)~\cite{smirnov}.  Here it is
approximated that $\theta_{13}=0$ and $\theta_{23}=\pi/4$. The
CP-asymmetry is
\be
\epsilon_1 \approx  -\frac {3A_1 M_1}{8\pi v^2} = 10^{-11} -10^{-10}
\( \frac{M_1}{10^6 \GeV}\).
\ee
The bound in Eq.~(\ref{eps}) is obtained for inverted hierarchy and
order one phases.  For quasi-degenerate neutrinos ($m_1 \approx m_2
\approx m_3 \gg (\Delta m^2_{\rm atm})^{1/2}$) the asymmetry is
suppressed by a factor $\Delta \sqrt{m_{\rm atm}^2}/\bar{m}$ with
$\bar{m}= 1/3\sqrt{m_1^2+m_2^2+m_3^2}$~\cite{Buchmuller}.

A similar calculation for decay of the next to lightest RH neutrino
gives
\bea
\epsilon_2 &=&
-\frac{1}{8\pi} \frac{1}{(h h^\dagger)_{22}}
\[ 
{\rm Im} \[ \{ (hh^\dagger)_{21} \}^2 \] \frac{M_1}{M_2}
+ {\rm Im} \[ \{ (hh^\dagger)_{23} \}^2 \] \frac{M_2}{M_3}
\] 
\nonumber \\
&\approx& 
-\frac{1}{8\pi} \[- \frac{m_u^2}{v^2} \frac{M_1}{M_2} + 3\frac{A_3}{v^2} M_2 \]
\nonumber \\
&\approx& 
2 \times 10^{-10} \( \frac{M_2}{10^6 \GeV}\)
\eea
where in the last step we have neglected the subdominant term
proportional to $M_1$, and we have used $A_3 < 6 \times 10^{-11} \GeV$
valid for hierarchical (light) neutrinos.  The bound is of the same
order of magnitude as the bound on $\epsilon_1$, but with $M_1
\leftrightarrow M_2$. Finally, for the CP-asymmetry of the heaviest RH
neutrino we get
\bea
\epsilon_3 &=&\frac{1}{8\pi v^2}
\[ \frac{M_1^2 A_1}{M_3} + \frac{M_2^2 A_2}{M_3} \]
\approx  \frac {A_2 M_2}{8\pi v^2}  \frac{M_2}{M_3}
\nonumber \\
&=& 3-7\times 10^{-11} \( \frac{M_2}{10^6 \GeV}\)\frac{M_2}{M_3}
\eea
where we have used $A_2 = 3 -5 \times 10^{-11}$ valid for
hierarchical light neutrinos.  Note that $\epsilon_3$ is suppressed
by a factor $M_2/M_3$ and is smaller than $\epsilon_1,\epsilon_2$.

\end{document}